Ice Giant Exploration Philosophy: Simple, Affordable

A white paper submitted to

The Committee on the Planetary Science Decadal Survey (2023-2032) of

The National Academies of Sciences

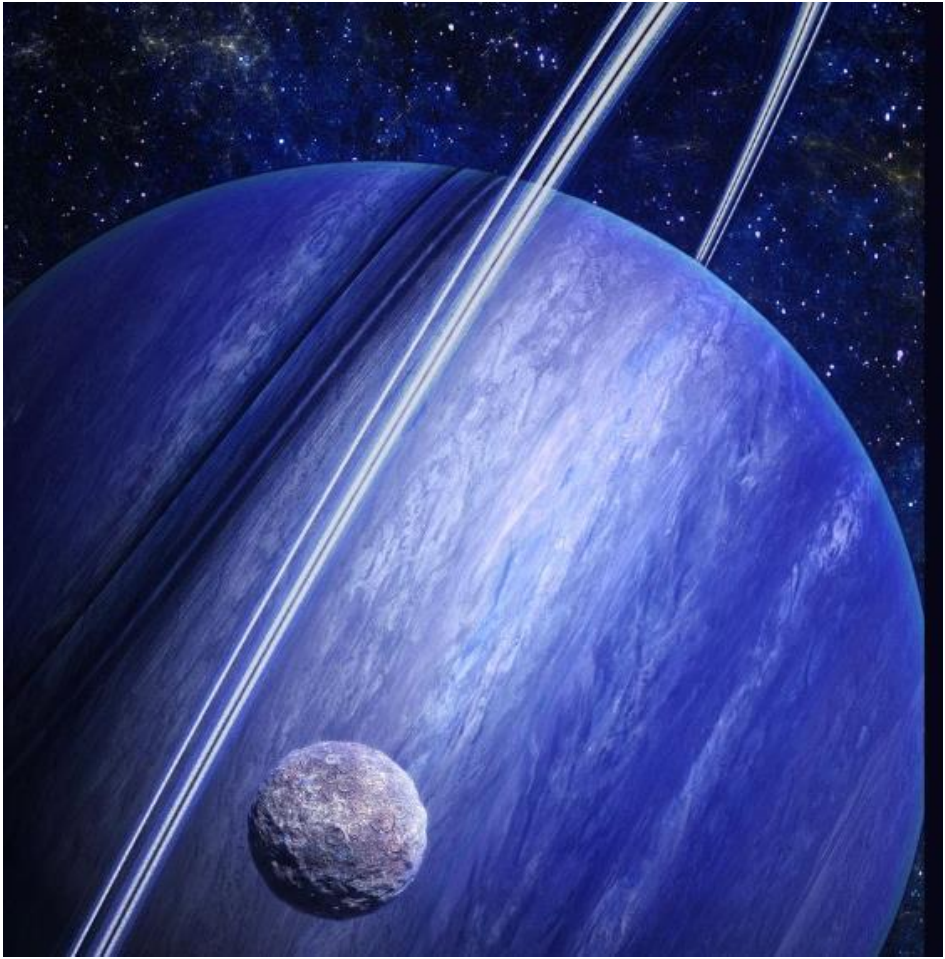

Philip Horzempa

LeMoyne College; Syracuse, New York

September 15, 2020

**Ice Giant Exploration Philosophy: Simple, Affordable**

For the Ice Giants, it is time to back away from the cutting edge of technology. Complexity produces delay and financial roadblocks. It is time to adopt the approach taken by the Messenger, Dawn and Io Observer teams. The science goals of an Ice Giant Flagship mission can be accomplished by a series of simpler, less-ambitious spacecraft over a period of years, instead of at one go.

This is a call for a radically different approach to any new missions to those worlds. The march of progress in spacecraft technology offers hope and a path forward. The key is to start small and keep it affordable.

**Introduction**

The Decadal Survey is asked to endorse a series of fast, simple, affordable (FSA) spacecraft to the Ice Giant systems. Separate lines of cost-capped Orbiters and Probes would be launched at a cadence dictated by trajectories and funding. Contractors would be selected using competitive Announcements of Opportunity.

The economic shadows of the 2020 Pandemic and Mars Sample Return make Flagship missions problematic. In addition, the uncertain cadence of RTG production dictates the use of solar power (where possible) and low-power avionics.

Fortunately, the march of progress has produced small spacecraft that can achieve the science goals of a Flagship mission. Power, propulsion, communications and instrument design have been revolutionized. This paper does not propose a specific mission design but, rather, asks the Decadal Survey team to endorse a philosophy of smaller, affordable missions to the Ice Giants.

NASA should allow private industry to propose their solutions. Missions such as Juno, Dragonfly and New Horizons demonstrate that they are up to the task. An AO will outline the science goals, schedule, and cost cap.

**Orbiters**

To begin, the Orbiters will be as small as possible. This translates into a very simple payload. This will constrain the complexity of these craft. The goal is to return to these Ice Giants for the first time in half a century. It will be a jaunt, so the lighter the load the better.

Crucial to success is the use of the private sector. NASA's New Frontiers, Discovery and Simplex missions have utilized that model resulting in low mission costs, less technical risk and timely execution.

Simple orbiters will be the first to be flown.  Keeping the exploration of the Ice Giant systems affordable translates into a successful tradeoff between mission scope and mission cost.  Some science goals will be deferred to later missions.  The first craft will concentrate on mapping and monitoring, setting the stage for the next orbiters in the series.  All of the science of a Flagship mission will be captured, but just not all at once.  It is the only way to make progress after half of a century of silence.

**A New Approach**:  If we are to survey the Ice Giant systems, then radical measures will need to be taken.   History offers several examples of how overly-ambitious projects gave way to ones that actually flew.

The first spacecraft to be named Voyager were grand spacecraft meant to land on Mars.  When NASA's budgets began to shrink in the late 1960s, the Voyager program was terminated.  It was replaced by a less-grand, but successful, Viking Mars lander project.

In the early 1970s, a Grand Tour program was in the works, meant to explore the outer planets.  The cost and complexity of the TOPS spacecraft led to its termination.  NASA scrambled to take advantage of the planetary alignment with a scaled-back Mariner-Jupiter-Saturn project, which inherited the Voyager name.  A mission extension for Mariner 12 (Voyager 2) allowed it to survey Uranus and Neptune for the first time.

Another example is the New Horizons mission.  It came about as the result of an impasse in the quest to launch a Pluto Express craft.  NASA studies produced designs that cost $2 billion.  As a result, the Express effort was canceled, leaving Pluto exploration in limbo.  In the year 2000 an AO was issued for such a mission, capping the cost at $1 billion.  The New Horizons proposal was chosen and the rest is history.

A similar impasse occurred with the Solar Probe project.  Costs for a Jupiter-flyby version had risen to $2 billion, resulting in an indefinite delay.  With a challenge to revamp the mission, NASA Goddard produced a $750 million design.  Its science goals were reduced, along with cost, but it won approval, with the result that the Parker Probe is now flying and returning data.

These examples demonstrate that productive missions can be flown within a fixed cost envelope.  Some science is sacrificed.  However, that is not quite a true picture.  Many complex missions returned no science data because they were too ambitious to be built and flown.  The limited-scope missions have succeeded because they moved from the world of fiction to that of hardware.

**Challenges**

**Timing**: Two launch windows that utilize Jupiter gravity assists will be available late in the decade.  In order to take advantage of the first Jupiter assist, it is imperative that Phase A should begin for a Neptune Orbiter in 2022.  This abbreviated timeline dictates the use of a simple craft, with no atmosphere Probe.

As it now stands, an Orbiter sent to Uranus to take advantage of a Jupiter gravity assist would not arrive until 2043, over 20 years from now.  A lightweight craft, however, could be sent on its way by 2029, arriving 7 years later.   Even though that is a half century after the flyby of Voyager 2, it would still be a full 17 years earlier than now envisioned.   Many in the planetary community can appreciate the advantage of this accelerated schedule.

The ability to create small, capable Orbiters means that both Ice Giants can be surveyed. There is no need to ignore one planet for the other.  Since so little is known about the Uranus and Neptune systems, it is pointless to prioritize one over the other.  Which one deserves a Flagship mission?  That is a question that is impossible to answer.  Therefore missions, limited in scope, should be sent to both targets.

**Budget**:  An examination of NASA's budget plan makes it clear that there will be stiff competition for funding in the 2020s.  A new Lunar Discovery program has been initiated that will absorb $5 billion in the coming decade.  In addition, the Mars Exploration line item is slated to increase substantially, while the planned outlays for Outer Planets are schedule to drop by 75%.  With Mars Sample Return likely to drain funds for the next 8-10 years, and with a dire national economic outlook, it could get worse.  How then are we to return to Uranus and Neptune?  A radically new approach is called for if we are to obtain any new data in the coming 20-40 years.

**Power Supply**:   In early 2020, the planetary community was informed that development of the enhanced-MMRTG had been halted.  There are plans for a Next-Gen RTG, but no guarantees that it will ever materialize.   Without the eMMRTG, all of NASA's Flagship designs have been rendered obsolete.  There are now only 2 options available, the standard MMRTG and solar power.  This limitation means that the Ice Giant craft will need to be very frugal with their power demands.

Considering the cadence of RPS production (and competition for their use from other NASA programs), it is probable that only one MMRTG unit will be available for an Ice Giant mission in the coming decade.  Two fast, simple, affordable (FSA) orbiters can be launched if one of those craft is solar-powered.  Physics dictates that that single MMRTG be used for the Neptune Orbiter.  In order to "get by" with one MMRTG, it will use power-saving electronics.  The Uranus orbiter can also "get by" with those same electronics, allowing it to fly with a solar array.

Advances in array technology mean that a Uranus orbiter can operate with solar power. The ROSA and Mega-ROSA panels can provide 200-400 W at 20 A.U.  The first ROSA array was launched to the ISS in 2017 and demonstrated its capability.

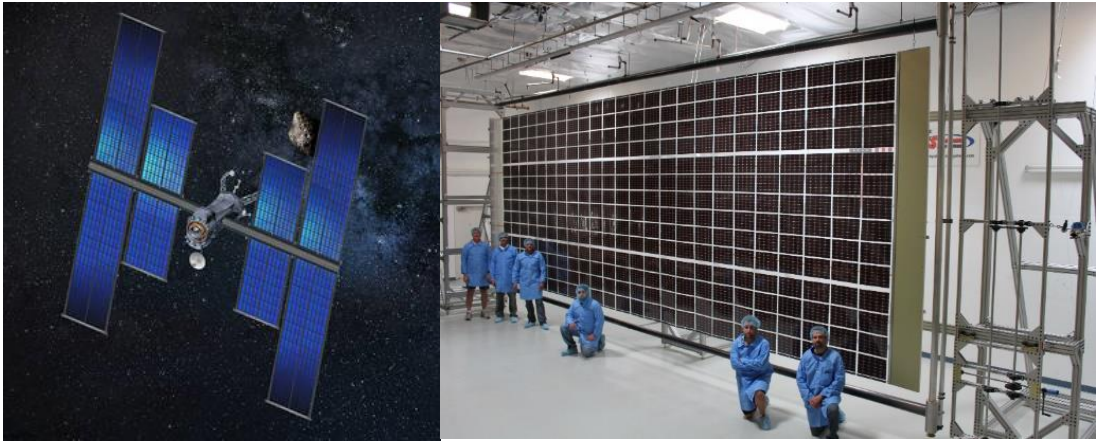

**Propulsion**: The Uranus Orbiter will be sent on a direct flight, without Jupiter GA, allowing a transit time as short as 7 years. The Neptune Orbiter, with Jupiter GA, has a transit time of 10 years. Heavy-lift rockets (Vulcan and Falcon) can send low-mass Orbiters on these fast-track trajectories.

Even though low in mass, these craft will need to decelerate into orbits around the Ice Giants from a high encounter velocity. Aerocapture would be ideal but it is a technology that does not yet exist. Another option is to use a disposable propulsion module. This could be modeled after ESA's Orbit Insertion Module to be used for the MSR Orbiter. An Ice Giant Orbiter would use the module for UOI and NOI, followed by ejection. The main orbiter would then utilize a less massive, less complex integrated propulsion system.

**Science Payload:** There are several approaches to limiting the mass and power-consumption of the instruments. One way is to limit the number of these devices. With a limited payload, first priority goes to imaging. The satellites of Uranus and Neptune are in dire need of complete, detailed photographic coverage. The first step in exploration is the creation of charts. It is a tradition that is thousands of years old. High-resolution and context cameras will produce those base maps. With an added near-IR capability, the imaging payload will also monitor the atmospheres of the Ice Giants and their ring systems.

A Vis-IR camera means that Uranus will be "transformed." Earth-bound near-IR images such as these show the atmospheric detail that will be available. An orbiter will produce images of Uranus' belts, zones and polar cyclones will resemble Juno's views of Jupiter.

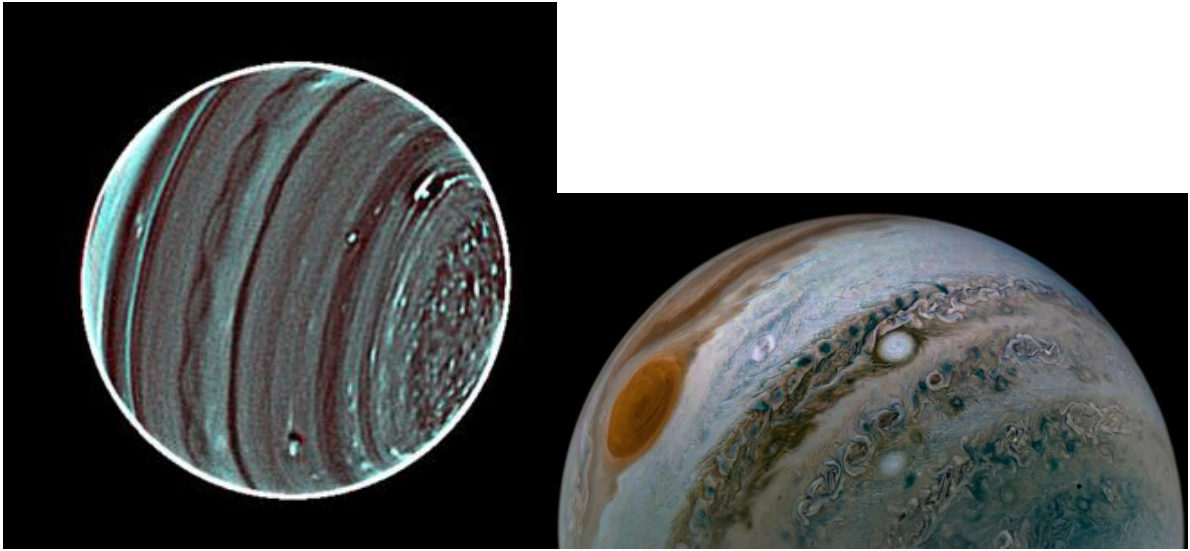

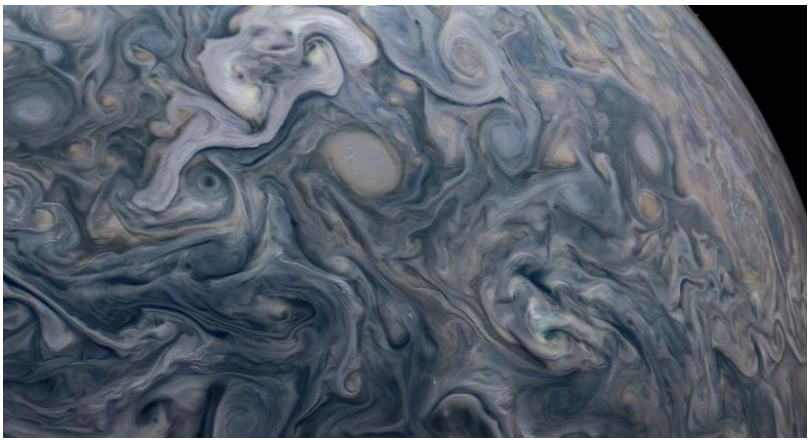

**Orbiters and Probes**

Key to affordability is the separation of the Probe missions from the Orbiters. This will allow the Orbiters to be simpler, cheaper. The Orbiters be given 1st priority in the launch queue. Since the Probe program will be untethered from the Orbiter effort, its mission cadence will be determined by factors unique to the study of Giant planet atmospheres.

**Orbiters**

The Ice Giant Orbiters will build on the experience of previous such missions. By now, industry has "figured out" how to construct such craft. This approach harkens back to the days of the first planetary orbiter, Mariner 9. It was an early example of a faster, cheaper, simpler project. Its main instrument was a camera, which produced the first global map of Mars. Simplicity allowed it to be made on-budget and on-time.

**Atmosphere Probes**

Industry has produced a number of options for low-cost, agile orbiter craft. The same cannot be said for Probes to the atmospheres of the Ice Giants. Probes should be sent to Uranus and Neptune, but technology readiness would be much more advanced now if a Saturn atmosphere Probe had paved the way. That mission has been on the docket for a good number of years, but several proposals by way of the New Frontiers program were rejected. Therefore, Probe development has stalled.

The Decadal Survey is asked to advocate for a separate Ice Giant Probe line. The initial mission would be a Saturn Probe. That would satisfy a long-standing objective and develop the technology required for almost-identical Probes for Uranus and Neptune. In addition, the Decadal Survey should advocate for combined KBO-Ice Giant Probe missions.

The Probe Carrier will also carry a Doppler Imager. Not only will it gather data on the interior of the planets, but it will provide images of the Probe entry site. If it is part of a KBO mission, such a camera will serve as the primary instrument during encounters with the dwarf worlds.

With the Orbiters no longer required to serve as Probe carriers, they can focus on their orbital reconnaissance of Uranus and Neptune.

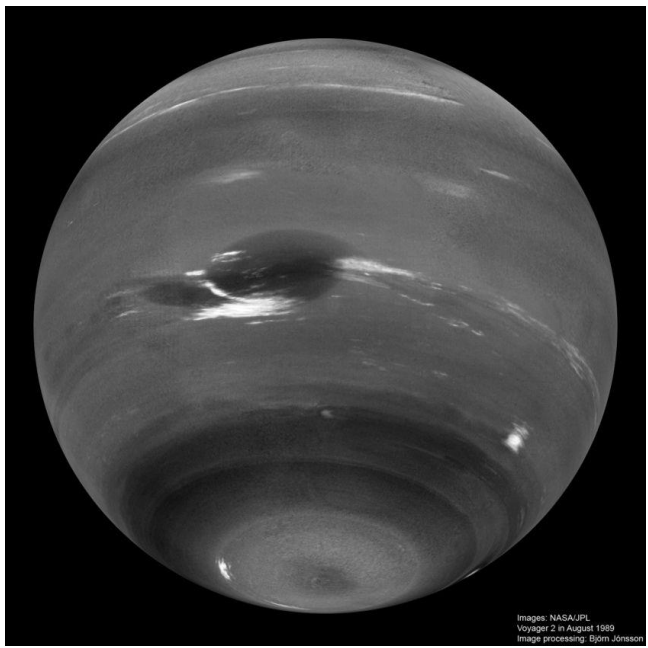

**Demands on NASA's Planetary Science Budget**

The exploration of the Ice Giant systems will not take place in isolation. There are other destinations in the solar system and scientists will want missions sent to their subjects of interest.

With the launch of the Mars 2020 rover, NASA is now committed to a Mars sample return program that will cast a financial shadow across the Planetary Science budget for the next 10 years. The MSR lander will be one of the most complex planetary vehicles ever launched. It must carry a Fetch Rover and a Mars Ascent Vehicle (MAV) to the surface. It needs to transfer samples to the MAV and serve as a launching pad for that rocket. The development, construction and testing will be a venture into new engineering territory. This mega-Flagship mission is not likely to leave the Earth before 2029, at a cost of $7 billion or more.

**Risks of Flagships and the Complexity Trap**

   Flagship missions are wonderful, but they are useless if they are so complex that they never get funded and never fly.  Less ambitious missions will deliver less science, but they have a better chance of achieving a coveted New Start.

**Conclusions**

   Ninety percent of life is showing up.  A simple, affordable series of spacecraft will allow us to "show up" at the Ice Giants for the first time in half a century.  The key is competition, akin to the philosophy of the New Frontiers, Discovery and Simplex programs.  Describe the mission, set the cost, and challenge the commercial sector to deliver.  The advent, and maturation, of Small Sat and Cube Sat capabilities since the last Decadal Survey allows science goals to be met with innovative, affordable technology.

   This paper does not put forward a specific design but, rather, asks the Decadal team to endorse a competitive approach to the exploration of the Ice Giant systems.  The key to rapid implementation will be the use of two separate series of spacecraft, Orbiters and Probes.  Their independence from each other will allow a more agile approach to funding and flying them.